\documentclass[11pt, a4paper]{article}

\usepackage[utf8]{inputenc}
\usepackage[T1]{fontenc}
\usepackage[english]{babel}
\usepackage[top=2.5cm, bottom=2.5cm, left=2.5cm, right=2.5cm]{geometry}
\usepackage{authblk}      
\usepackage{graphicx}     
\usepackage{booktabs}     
\usepackage{longtable}    
\usepackage{array}
\usepackage{url}
\usepackage{hyperref}
\usepackage{enumitem}
\usepackage{float}
\usepackage{listings}
\usepackage{xcolor}
\usepackage{caption}

\hypersetup{
    pdftitle={Cortex: Content Analysis Support Software, a Resource for Qualitative Research},
    pdfauthor={Ana Julia da Silva Soares},
    pdfsubject={Computer Science - Software Engineering},
    pdfkeywords={Content Analysis, Qualitative Research, Artificial Intelligence, NLP, RAG},
    colorlinks=true,
    linkcolor=blue,
    citecolor=blue,
    urlcolor=blue
}

\title{\textbf{Cortex: Content Analysis Support Software, a Resource for Qualitative Research}}

\author{Ana Julia da Silva Soares}
\author{Rafael Coimbra Pinto}
\affil{Instituto Federal de Educação, Ciência e Tecnologia do Rio Grande do Sul (IFRS), Campus Canoas, Brazil}
\date{2026}

\begin{document}

\maketitle

\begin{abstract}
Qualitative research is widely used in the human and social sciences, characterized by a deep understanding of phenomena through the interpretation of meanings and contexts. Among qualitative data analysis methods, content analysis stands out as a consolidated technique, which allows for the systematic description and interpretation of textual contents. However, as data volume increases, the time required for organization, reading, and categorization becomes a significant challenge, potentially delaying research development. Therefore, this work aimed to develop a web application to support content analysis, based on Bardin's methodology, targeted at academic researchers. The methodology adopted a mixed approach, combining bibliographic research on content analysis with semi-structured interviews with four experienced researchers, aiming to identify real needs and requirements. Based on these inputs, the Cortex software was developed to assist the researcher in the pre-analysis and material exploration stages, generating suggestions for indices, indicators, and categories with full traceability to original documents. The results demonstrate that Cortex is capable of guiding the researcher through the entire methodological workflow, from corpus configuration to results exportation, acting as a methodological collaborator without replacing the researcher's interpretative autonomy.

\vspace{0.5cm}
\end{abstract}

\section{Introduction}

Qualitative research is a methodological approach widely used in human and social sciences, characterized by its focus on the deep understanding of social, cultural, and subjective phenomena \cite{breuer2002}. According to Godoy \cite{godoy1995}, this type of research is not based on the quantification of events or the application of statistical instruments, but starts from broad questions that are delineated throughout the study. The author highlights that such research seeks to describe people, contexts, and processes based on the direct interaction between the researcher and the object of study, prizing the understanding of phenomena from the perspective of the subjects involved \cite{godoy1995}. Unlike quantitative approaches designed to measure variables, qualitative studies rely on the interpretation of meanings, contexts, and experiences, employing methods such as interviews, observations, and discourse analysis \cite{minayo2001}.

Among the methods used to analyze textual data obtained in this domain, Content Analysis stands out as a consolidated and effective technique originating in the late 20th century. According to Moraes \cite{moraes1999}, it is a methodology that allows describing and interpreting the contents of different types of documents and texts, enabling a deep understanding of the core messages. When applied with rigor, this approach transforms raw data into structured knowledge, being widely recognized for its theoretical and methodological consistency \cite{sousa2020}.

However, as the volume of data increases — especially in contemporary studies involving multiple dense interviews or large-scale open-ended responses — the time and cognitive effort required for organizing, reading, and analyzing content become significant challenges \cite{sousa2020}. Coder fatigue, tight deadlines, and the inherent complexity of manual coding rules can compromise analysis precision and introduce subjective biases \cite{sampaio2018}. Bardin \cite{bardin2011} emphasizes that maintaining an unbiased and systematic analytical stance demands a continuous psychological effort from the analyst, stating that:

\begin{quote}
"structural deciphering centered on each interview [requires] an effort but that does not exclude intuition, to the extent that, in each new interview, it is necessary to make an abstraction of oneself and previous interviews [...], to make a clean slate, a priori, of personal opinions or contamination from previous decipherings" \cite[p. 96]{bardin2011}.
\end{quote}

Considering that qualitative analysis is only one
stage of the research process, the time spent in this phase can delay the general development of
the study. Therefore, tools that contribute to speeding up this stage, without compromising the
quality of the analysis, can represent an important support to scientific activity.

To address these limitations, this work presents Cortex, a web application designed to support academic researchers throughout the explicit chronological phases of Content Analysis, according to the methodology proposed by Bardin \cite{bardin2011}. The tool allows users to import their textual research corpus and theoretical frameworks to automatically generate suggestions for indices, indicators, and thematic categories via a Large Language Model (LLM), while maintaining strict verification and traceability to the original source text.

\section{Methodology and System Design}

This work adopts a mixed approach, combining quantitative and qualitative methods, as suggested by Creswell and Plano Clark \cite{creswell2011}, focusing on the development of a web application to support researchers in the qualitative content analysis process.

Initially, a bibliographic search was carried out on content analysis methodology and its applications in qualitative research to theoretically base the tool's development. In parallel, four semi-structured interviews were conducted with researchers who have experience with content analysis methodology or qualitative methods in general.

The group of participants consisted of professors and researchers linked to higher education institutions, all with postgraduate degrees (master's and doctorate). The multidisciplinarity of the group, covering distinct areas such as Computer Science, International Relations, and Education, was instrumental in capturing diverse cross-domain challenges.

The objective of this stage was to identify needs, difficulties, and expectations regarding the use of digital tools in the process of categorization, visualization, and analysis of textual data. The interviews were transcribed and analyzed qualitatively, allowing the identification of requirements aligned with the real demands of future application users.

Based on the requirements and inputs obtained in this initial phase, the system was modeled to graphically represent the main functionalities and structural entities. The database modeling and the domain classes of Cortex were designed to reflect the Content Analysis methodology proposed by Bardin \cite{bardin2011}. Each methodological stage is represented by specific entities that store the artifacts generated in each phase. Consequently, the core structure includes strictly defined entities for the Analysis, the Document, the Index, the Indicator, the Category, and the Register Unit.

Once modeling was completed and requirements defined, the software development phase began. To operationalize this theoretical framework, the software architecture was divided into four cohesive functional modules: Authentication, Configuration, Pre-Analysis, and Exploration. 

\begin{figure}[H]
    \centering
    \includegraphics[width=0.6\textwidth]{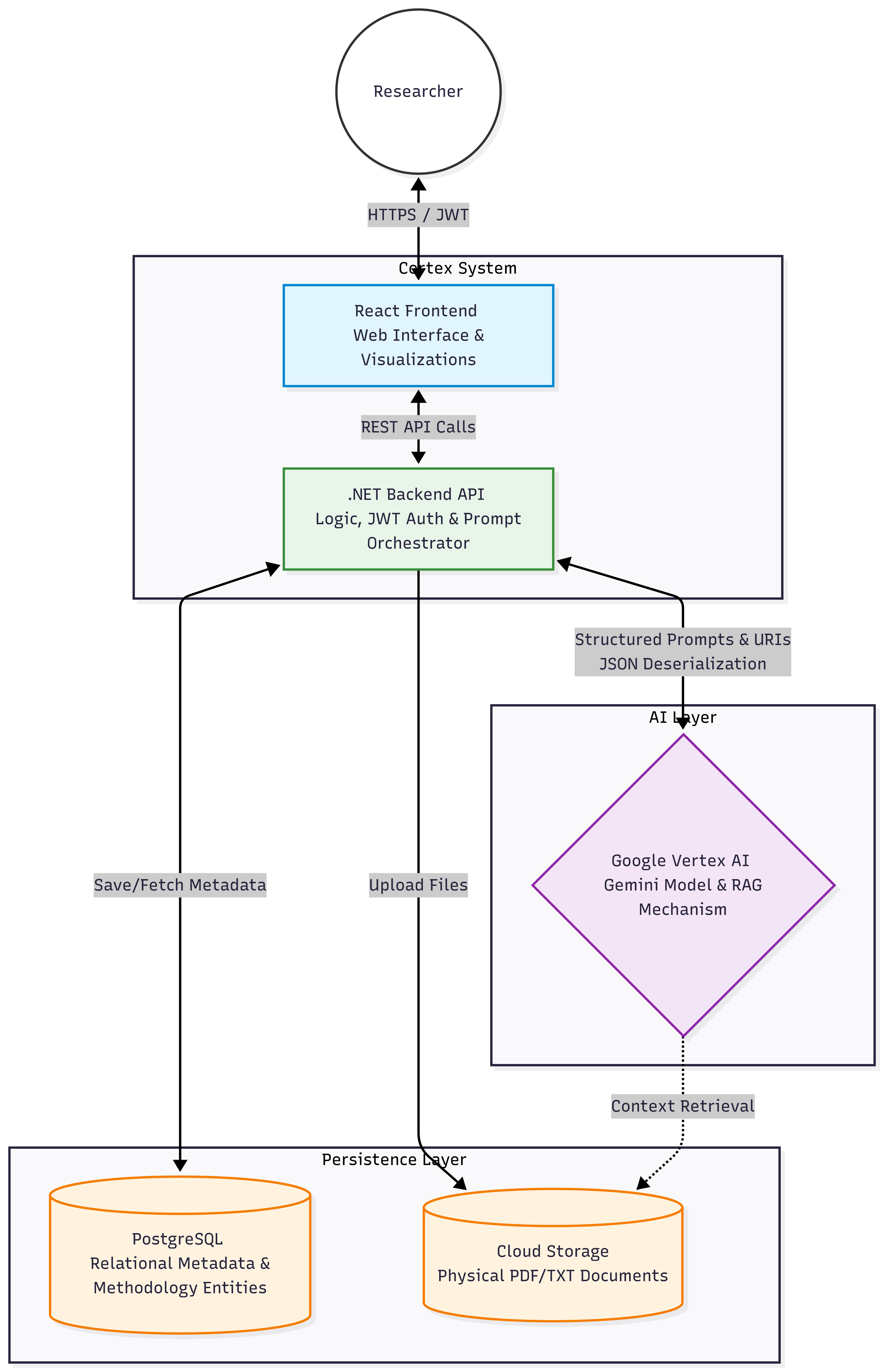} 
    \caption{High-level architecture of the Cortex system, illustrating the interaction between the client interface, the backend API, hybrid persistence layers, and the LLM RAG mechanism.}
    \label{fig:architecture}
\end{figure}

The application was built using a decoupled client-server model, utilizing C\# with .NET in the backend to provide a robust, high-performance API. The frontend was developed using React with TypeScript, ensuring a responsive user interface capable of managing dynamic data visualizations. The structural organization and data flow between these components are illustrated in Figure \ref{fig:architecture}.

To elucidate the dynamic behavior of Cortex, the data flow adopts a hybrid storage strategy during the configuration stage, which corresponds to the constitution of the corpus. Physical files, such as PDFs or text documents, are sent directly to Google Cloud Storage to guarantee scalability, while the local PostgreSQL database stores only the metadata and the logical references, or URIs, returned by the cloud. 

For the intelligent processing phases, the system integrates the Gemini model through its API on the Vertex AI platform, applying the Retrieval-Augmented Generation (RAG) technique. In both the Pre-Analysis and Exploration stages, the backend queries the database for the document URIs, elaborates a methodological prompt, and defines a strict JSON Schema. These inputs are sent to Vertex AI, which reads the original documents and generates a structured response using Natural Language Processing (NLP) techniques \cite{zhou2020}. Finally, the LLM's response is validated by the schema, automatically deserialized into C\# domain classes, and persisted in the relational database, allowing the frontend to render the final results in comprehensive tables and graphs.

\section{Theoretical Background}

\subsection{Qualitative Research and Content Analysis}
Qualitative research is a methodology that is not concerned with numerical representativeness, but with deepening the understanding of a social group, an organization, or phenomena that cannot be quantified, focusing on explaining the dynamics of social relations \cite{gerhardt2009}. According to Godoy \cite{godoy1995}, qualitative research does not seek to enumerate or measure events, nor does it employ statistical instruments in data analysis, but rather starts from broad questions that are defined throughout the study.

Content analysis aligns directly with qualitative research, defined as a methodology of this nature, especially in the field of Social Sciences \cite{cardoso2021}, but which can also be used to analyze quantitative data, according to Bardin \cite{bardin2011}. This approach goes beyond the simple description of qualitative data, allowing new interpretations that deepen the understanding of underlying meanings present in collected information \cite{moraes1999}.

Because of this deeply interpretative nature, Minayo \cite{minayo2001} notes that content analysis inevitably involves the researcher's personal lens. She argues that a completely neutral, objective, and exhaustive reading is unattainable, as the values, language, and cultural context of both the researcher and the object of study actively influence data interpretation. 

To mitigate this inherent subjectivity and ensure scientific rigor, the methodology must follow a highly structured workflow. In this regard, Bardin \cite{bardin2011} organizes the phases of Content Analysis around three explicit chronological poles: pre-analysis; exploration of the material; and treatment of results, inference, and interpretation.

\subsubsection{Pre-analysis}
This is the initial phase of organization, intuition, and systematization of ideas. In this phase (also called corpus constitution), the choice of documents to be analyzed is made. According to Bardin \cite{bardin2011}, the first sub-stage is "floating reading", where the researcher comes into contact with the material.

Still in this stage, preparation for future data interpretation occurs through the elaboration of indicators. Bardin describes the corpus as a manifestation carrying 'vestiges' (indices) that the analysis must 'make speak'. The researcher's task is to identify these indices on the textual surface and organize them systematically into indicators. While indices are concrete signs in the text, indicators represent the measurement rule or method adopted to quantify (or qualify) these signs \cite{cardoso2021}.

To illustrate this distinction, Bardin \cite{bardin2011} uses a classic example of a therapeutic interview. The author proposes a hypothesis where emotion and anxiety (the concept to be inferred) would manifest through speech disturbances. In this scenario, the indices (vestiges) are the concrete signs in communication, such as interrupted sentences, stuttering, or incoherent sounds (e.g., "uh", "um"). The indicator (measure), in turn, is the counting rule defined by the researcher, such as the frequency of appearance of these indices. It is this transformation of an abstract phenomenon into a concrete indicator that allows the analysis to be objective and systematic. Table \ref{tab:indices_indicators} summarizes this relationship.

\begin{table}[htbp]
\centering
\caption{Example of the relationship between hypothesis, indices, and indicators in content analysis}
\label{tab:indices_indicators}
\begin{tabular}{|p{4cm}|p{4cm}|p{4cm}|p{3cm}|}
\hline
\textbf{Concept to be Inferred (Inferred Variable)} & \textbf{Hypothesis} & \textbf{Indices (Vestiges in Text)} & \textbf{Indicator (Systematic Measure)} \\ \hline
Anxiety/Emotion & Anxiety manifests through speech disturbances & Interrupted sentences, word repetition, stuttering, incoherent sounds (e.g., "uh", "um") & The frequency of appearance of these indices \\ \hline
\end{tabular}
\small \\ Source: Elaborated by the author, based on Bardin (2011).
\end{table}

\subsubsection{Exploration of the Material}
The material exploration stage consists of the systematic application of the decisions made during pre-analysis. According to Bardin \cite{bardin2011}, this is the phase where the collected material is effectively transformed into analyzable data through coding operations. With the objectives, corpus, and indicators already defined, the researcher applies the analysis techniques to the text.

The central operation of this phase is material coding. Before classifying excerpts from the corpus, the text must be broken down into smaller, manageable segments. The most common technique, Categorical Analysis (or Thematic Analysis), begins by dividing the text into two distinct units: the recording unit and the context unit \cite{bardin2011}. 

The recording unit is the core unit of meaning (e.g., a specific theme or word) that the researcher will register and count. Conversely, the context unit is the larger text segment surrounding the recording unit, providing the necessary background to fully comprehend its exact meaning. Context is crucial in qualitative analysis, as the true significance of isolated elements heavily depends on the broader discourse \cite{bardin2011}.

After segmenting the text, the analyst groups these units into categories. Categories act as conceptual "drawers" or significant rubrics that allow the classification of the message's constitutive meaning elements. Once categorization is established, the researcher's task is to count the frequency of presence (or absence) of these meaning items within each category. In Categorical Analysis, the frequency with which certain content features appear serves as the primary analytical information.

Given the highly systematic nature of this phase—which involves meticulous coding, categorization, and frequency counting—Bardin \cite{bardin2011} highlighted that the use of computers is a remarkably valuable resource. The author emphasized their utility for complex analyses involving multiple variables, occurrence tracking, and elaborated statistical operations, significantly accelerating a process that would otherwise be exhaustingly manual. This precise methodological gap is the primary target of the Cortex system's automation capabilities.

\subsubsection{Treatment of Results, Inference, and Interpretation}

In this phase, raw data is processed to highlight and interpret the obtained information \cite{gerhardt2009}. The results can be condensed and visually represented through diagrams, figures, and models \cite{cardoso2021}.

Based on these results, inferences can be made, which represent the essential purpose of this methodology. As Bardin explains, "the intention of content analysis is the inference of knowledge related to the conditions of production (or, eventually, reception), an inference that relies on indicators (quantitative or otherwise)" \cite[p. 44]{bardin2011}.

Interpretation (the assignment of meaning to the detected characteristics) is the final stage, while inference serves as the intermediate procedure that allows an explicit and controlled transition from the description phase to the interpretation \cite{bardin2011}. 

This link between objective description (the counted data) and interpretation (the assigned meaning) is bridged by the theoretical framework. Therefore, inference is not random; it consists of analyzing the descriptive results in light of the hypotheses and theories that grounded the pre-analysis. It is this return to theory that allows the researcher to propose meanings for the findings, going beyond what is merely explicit in the text \cite{bardin2011}.

Finally, the validity of Content Analysis should not be judged against a single "true reading" of the text — since the meaning of a text is rarely entirely manifest or unique — but rather by its grounding in the researched materials, its congruence with the researcher's theoretical framework, and its alignment with the research objective \cite{cardoso2021}.

\subsection{Natural Language Processing and Artificial Intelligence}
Natural Language Processing (NLP) is a subfield of artificial intelligence focused on enabling computers to understand and process human languages \cite{howard2019}. It studies fundamental technologies for expressing word, phrase, sentence, and document meanings. The most significant advances in NLP in recent decades have been driven by AI, particularly Machine Learning (ML) and Deep Learning.

\subsection{Retrieval-Augmented Generation (RAG)}
While Content Analysis is a human inference methodological process, applying computational NLP techniques allows automating parts of this process. Large Language Models (LLMs) have become prominent, but they have limitations such as hallucinations and knowledge cutoffs \cite{james2025}.

To solve these limitations, Retrieval-Augmented Generation (RAG) architecture was developed. This technique transforms the LLM from a closed knowledge agent to a system that reasons based on external and updated information sources. It consists fundamentally of two phases: Retrieval and Generation.

For the RAG system to work, three technical components are fundamental: Embeddings (Vectorization), Chunking (Fragmentation), and a Similarity Search mechanism. The process begins with chunking, which divides extensive documents into smaller, semantically cohesive fragments. Then, embeddings are used to create a numerical representation of this data in a high-dimensional vector space. Finally, similarity retrieval compares the user query vector with the chunk vectors to find the relevant context.

\subsubsection{Vertex AI}
The implementation of a RAG system involves the orchestration of all the aforementioned components: document preprocessing, chunking, embedding generation, vector indexing, and the call to the LLM \cite{james2025}. Cloud platforms, such as Google Cloud's Vertex AI, offer managed services that abstract this complexity.

According to the official documentation \cite{google2024}, the Vertex AI platform allows users to provide their source documents (such as PDFs and TXTs). The tool then automatically manages the data ingestion pipeline, which includes parsing, chunking, and indexing (embedding creation) the data. Once indexed, the corpus becomes available to be queried by the Gemini model, which utilizes the platform's RAG mechanism to retrieve relevant excerpts and ground its responses \cite{google2024}. Therefore, utilizing this API enables the application of the RAG technique without the need to manually build and maintain each component of the pipeline.

\section{Related Work}

\subsection{ATLAS.ti}
ATLAS.ti is one of the most consolidated qualitative analysis software in academia. Its structure is influenced by Grounded Theory. The workflow relies on creating a "project" where the researcher imports documents and performs "coding" (selecting text excerpts and assigning codes). Its strength lies in managing and visualizing relationships between codes (semantic networks). Recently, it incorporated LLM-powered features, but its core remains a sophisticated "workbench" for manual process \cite{muhr1991}.

\subsection{NVivo}
NVivo is the main competitor to ATLAS.ti. It allows coding text excerpts into "Nodes". Its query resources are a differentiator, allowing complex searches and coding matrices. Like its competitor, it has introduced LLM modules, but requires the researcher to read, interpret, and code material manually before inferring \cite{qsr2014}.

\subsection{Elicit}
Elicit represents a paradigm shift. It is not a traditional CAQDAS, but an "AI research assistant" native to the LLM era. Its main focus is automating research tasks, especially literature review. Using RAG, Elicit allows users to ask questions in natural language, and the tool locates scientific articles and extracts key information. However, its design is not aimed at analyzing primary data (like interviews) under a specific qualitative methodology \cite{elicit2024}.

\subsection{Comparative Analysis and Gap Identification}

The analysis of the tools demonstrates clear differences. On one hand, ATLAS.ti and NVivo offer robust environments for manual methodological rigor, excelling at managing the complexity of coding, but they have a steep learning curve and do not automate the cognitive phases of the analysis. On the other hand, Elicit uses advanced LLMs to automate synthesis but focuses on literature review and does not adapt to the prescriptive steps of methodologies such as Bardin's \cite{bardin2011}.

None of the analyzed tools were designed to fill the central gap identified by this work: the application of Large Language Models (integrated with the RAG architecture) to actively assist the researcher within the specific phases of Content Analysis.

While NVivo and ATLAS.ti treat the researcher as the sole agent of inference, and Elicit acts as an automated literature synthesizer, the proposal of this work (the Cortex software) positions itself as a methodological collaborator. The objective of Cortex is to use a Large Language Model via the Gemini API not to replace the researcher, but to accelerate the pre-analysis and exploration stages by suggesting preliminary indices, indicators, and categories based on the provided corpus and theoretical framework. Table \ref{tab:comparative_analysis} below summarizes the main differences between the analyzed tools and the proposal of this work.

\begin{table}[htbp]
\centering
\caption{Comparative analysis of data analysis tools}
\label{tab:comparative_analysis}
\begin{tabular}{|p{3cm}|p{4cm}|p{3cm}|p{4cm}|}
\hline
\textbf{Feature} & \textbf{ATLAS.ti / NVivo (CAQDAS)} & \textbf{Elicit (AI-Native)} & \textbf{Cortex (Proposed)} \\ \hline
\textbf{Main Methodology} & CAQDAS (Grounded Theory, Thematic Analysis, etc.) & Generative AI (LLMs) & Content Analysis (Bardin) assisted by AI (RAG) \\ \hline
\textbf{Usage Focus} & Primary data analysis (interviews, audio, etc.) & Secondary data analysis (literature review) & Primary data analysis (interviews, documents) \\ \hline
\textbf{Paradigm} & Manual and Assisted: manager of manual codings & Automated: task executor & Collaborative and Assisted: suggestion of indices/categories based on RAG \\ \hline
\textbf{Alignment with Bardin} & Indirect: allows manual application & Non-existent & Direct: designed to follow Bardin's steps \\ \hline
\end{tabular}
\end{table}

\section{Results: The Cortex Software}

The development of Cortex resulted in a robust web application that actively guides the qualitative researcher through the chronological phases of Content Analysis. The system's interface and underlying logic were designed to abstract the technical complexity of Large Language Models (LLMs), presenting a workflow entirely focused on methodological rigor.

\subsection{Authentication and Workspace Initialization}
The researcher's interaction begins in a secure, authenticated environment utilizing JSON Web Tokens (JWT) to ensure strict data isolation between users and projects. Upon authentication, the main dashboard provides a tabular overview of all ongoing analyses. Given the extensive time required for qualitative research, the system implements continuous persistence, allowing the user to pause their work at any methodological stage and resume later without data loss. To initiate a new project, the researcher simply defines the analysis title, immediately advancing to the corpus constitution phase.

\subsection{Configuration Stage: Corpus Constitution}
The configuration stage directly mirrors the initial corpus constitution described by Bardin \cite{bardin2011}. In this environment, the researcher establishes three fundamental elements. First, they must define the central research question, which acts as the primary semantic anchor for all subsequent LLM processing. Following this, the user uploads the textual documents to be analyzed, typically interview transcripts in PDF or TXT formats, which form the primary corpus. Finally, the system allows the optional upload of reference documents, such as scientific articles or theoretical frameworks. These reference files are not analyzed as primary data but are ingested by the RAG mechanism to contextualize and ground the LLM's future suggestions. Once these parameters are validated, the system strictly enforces linear progression, preventing backward alterations to maintain methodological integrity.

\subsection{Pre-Analysis: LLM-Driven Index Generation}
Upon advancing, Cortex initiates the Pre-Analysis stage by orchestrating the first interaction with the Gemini model. The backend constructs a highly structured prompt incorporating the central research question, the corpus metadata, and the explicit criteria defined by Bardin \cite{bardin2011} for index selection and indicator construction. This prompt, along with the document URIs, is processed by Vertex AI using the RAG architecture. 

\begin{figure}[H]
    \centering
    \includegraphics[width=0.9\textwidth]{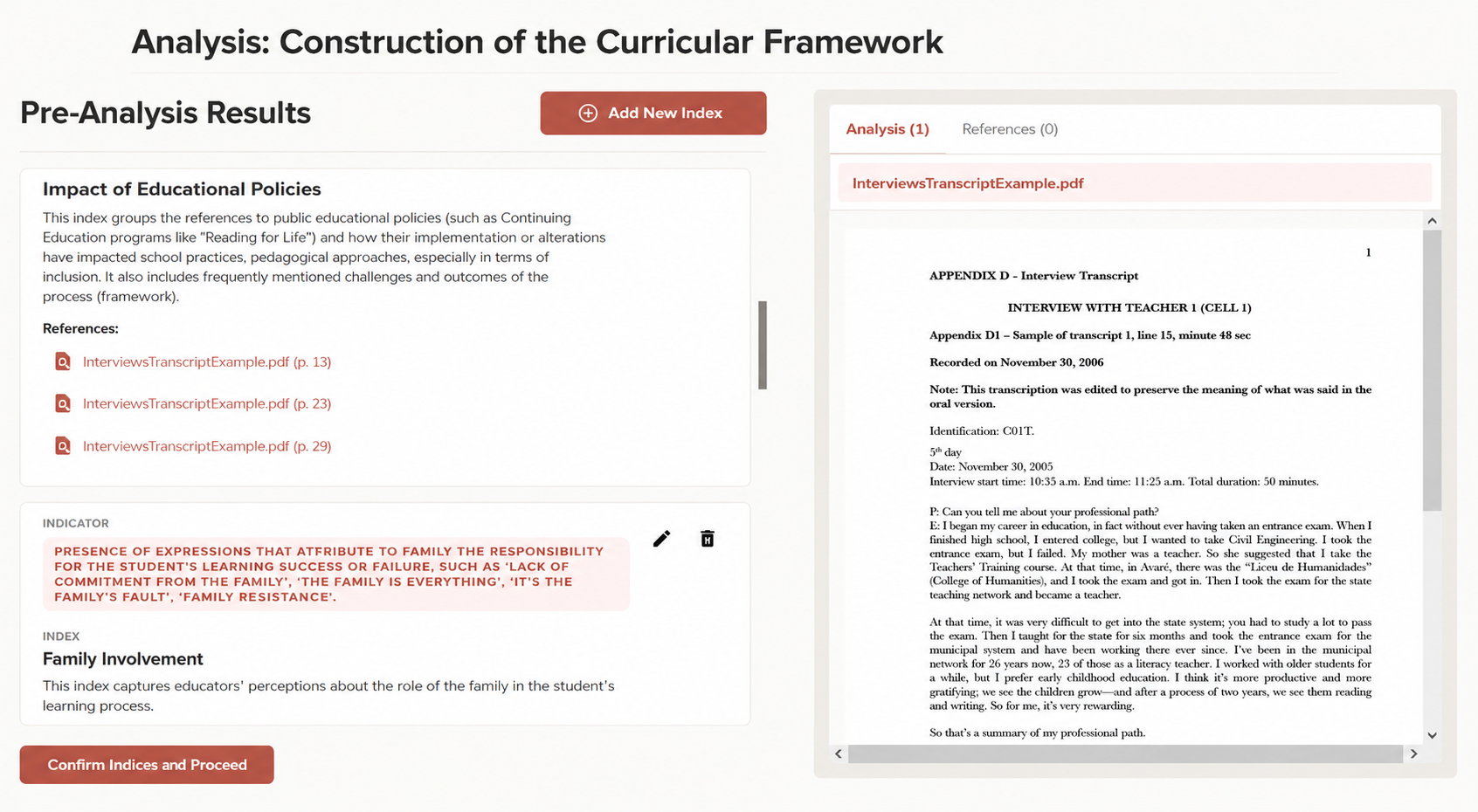} 
    \caption{The Pre-Analysis interface displaying LLM-generated indices alongside the integrated document viewer. Key visual elements include the suggested index name and description, the systematic counting rule, and the exact textual extractions linking the LLM's conceptual suggestion directly to the source document and page number.}
    \label{fig:preanalysis}
\end{figure}

The LLM's response is deserialized and presented to the researcher as interactive cards, as illustrated in Figure \ref{fig:preanalysis}. Each card displays a suggested index, its theoretical description, the associated indicator, and a critical list of exact textual references extracted directly from the uploaded documents, complete with page numbers. To enhance traceability, an integrated document viewer positioned alongside the cards allows the researcher to simultaneously read the full original context of any generated reference in real-time, validating the LLM's inferences against the raw data.

\subsection{Human-in-the-Loop: Concept Appropriation}
A critical design decision in Cortex is the enforcement of researcher autonomy over the LLM's outputs. During the Pre-Analysis stage, the researcher exercises full methodological appropriation by editing the theoretical names and descriptions of the generated indices, removing irrelevant suggestions, or creating entirely new indices manually. However, the system deliberately prevents the user from editing the raw textual references and page numbers extracted by the LLM. This constraint guarantees that while the theoretical interpretation remains entirely under human control, the documentary foundation remains immutable, ensuring absolute analytical traceability.

\subsection{Material Exploration and Automated Categorization}
The Material Exploration stage represents the most technically complex phase of the system, where the thematic categorization of the corpus occurs. The system retrieves all the indices and indicators previously validated by the researcher and constructs a second structured prompt. This new instruction commands the LLM to identify specific register units within the text and group them into cohesive categories based strictly on the validated indices and frequency counting rules. 

\begin{figure}[H]
    \centering
    \includegraphics[width=0.9\textwidth]{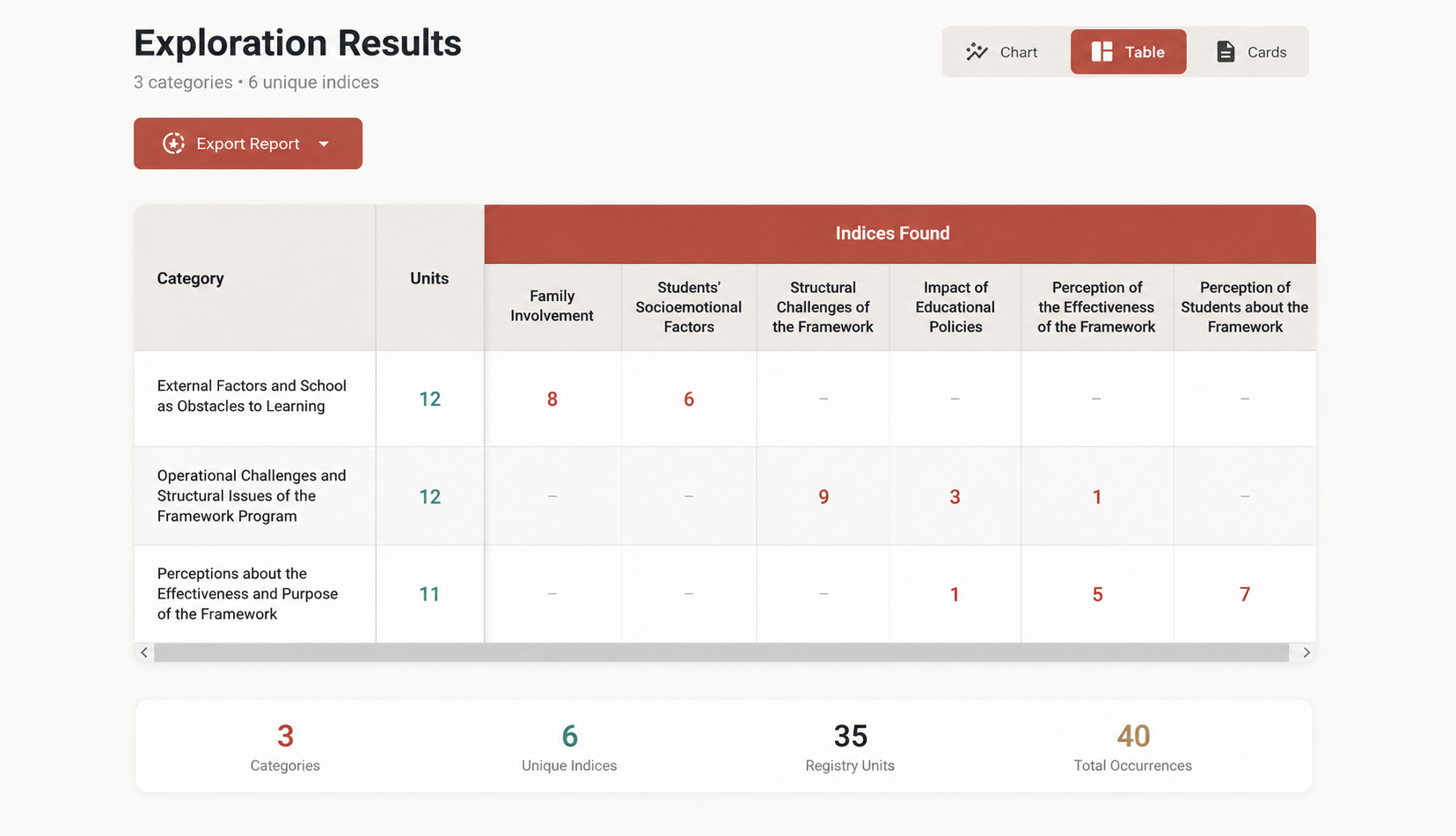} 
    \caption{The Material Exploration interface displaying the categorization matrix. The table illustrates the final methodological output, cross-referencing the generated Categories, their total frequencies, and the specific distribution of Found Indices within each thematic group.}
    \label{fig:exploration}
\end{figure}

The results of this exploration are rendered in three complementary visualizations. First, a graphical bar chart displays the frequency distribution of the categories, allowing for a rapid visual assessment of thematic representativeness. Second, a tabular categorization matrix, as presented in Figure \ref{fig:exploration}, details the co-occurrence frequencies between the identified categories and the applied indices. Finally, an expandable card layout presents each category's semantic definition, total frequency, and the exhaustive list of related register units. Each register unit displays the exact textual excerpt, its source document, and a justification for its categorization. 

Notably, unlike the Pre-Analysis stage, these final categories are non-editable. This architectural restriction ensures methodological validity, as the categories must remain the direct, systematic mathematical result of applying the previously validated indices over the corpus.

\subsection{Results Export and Methodological Traceability}
To conclude the analytical workflow, Cortex generates a comprehensive PDF report containing the entire analytical trail. This final artifact encompasses the project's metadata, the central research question, the frequency distribution graphs, and the complete categorization matrix. Furthermore, it extensively details every category, its associated indices, the employed indicators, and an exhaustive list of all register units with their exact textual origins and justifications. This exported report serves as a definitive, traceable record, enabling readers and peer reviewers to follow the exact methodological path from the raw textual corpus to the final conceptual inferences.

\section{Final Considerations}

This work developed a web application to support content analysis, grounded in Bardin's methodology \footnote{The source code and repository for the Cortex project are publicly available at \url{https://github.com/AnaJuliaSSdev/Cortex}.}. Cortex acts as a methodological assistant, automating laborious steps (index identification, counting) while preserving human control over interpretative decisions.
The development process followed specific objectives, from theoretical research to implementation using modern stack (.NET, React, Vertex AI).
Results show the system successfully guides researchers through the workflow. Limitations include export restricted to PDF in the MVP. Future work includes multi-format export, audio transcription integration, and advanced visualizations (word clouds).

\bibliographystyle{plain}

\begin{thebibliography}{99}

\bibitem{bardin2011}
Bardin, L. (2011). \textit{Análise de conteúdo}. São Paulo: Edições 70.

\bibitem{breuer2002}
Breuer, F., \& Reichertz, J. (2002). Standards of social research. \textit{Forum: Qualitative Social Research}, 2(3).

\bibitem{cardoso2021}
Cardoso, M. R. G., Oliveira, G. S., \& Ghelli, K. G. M. (2021). Análise de conteúdo: uma metodologia de pesquisa qualitativa. \textit{Cadernos da FUCAMP}, 20(43), 92-106.

\bibitem{creswell2011}
Creswell, J. W., \& Plano Clark, V. L. (2011). \textit{Designing and Conducting Mixed Methods Research}. 2nd ed. Thousand Oaks: SAGE.

\bibitem{elicit2024}
Elicit. (2024). \textit{Elicit: The AI Research Assistant}. Available at: https://elicit.com/.

\bibitem{gerhardt2009}
Gerhardt, T. E., \& Silveira, D. T. (orgs.). (2009). \textit{Métodos de pesquisa}. Porto Alegre: Editora da UFRGS.

\bibitem{godoy1995}
Godoy, A. S. (1995). Introdução à pesquisa qualitativa e suas possibilidades. \textit{RAE - Revista de Administração de Empresas}, 35(2), 57-63.

\bibitem{google2024}
Google. (2024). \textit{Retrieval-Augmented Generation (RAG) Overview}. Google Cloud Vertex AI.

\bibitem{howard2019}
Howard, J. (2019). Artificial intelligence: Implications for the future of work. \textit{American Journal of Industrial Medicine}, 62(11), 1-10.

\bibitem{james2025}
James, A., Trovati, M., \& Bolton, S. (2025). Retrieval-Augmented Generation to Generate Knowledge Assets. \textit{Applied Sciences}, 15, 6247.

\bibitem{minayo2001}
Minayo, M. C. S. (2001). \textit{Pesquisa social: teoria, método e criatividade}. 18th ed. Petrópolis: Vozes.

\bibitem{moraes1999}
Moraes, R. (1999). Análise de conteúdo. \textit{Revista Educação}, 22(37), 7-32.

\bibitem{muhr1991}
Muhr, T. (1991). ATLAS.ti: a prototype for the support of text interpretation. \textit{Qualitative Sociology}, 14(4), 349-371.

\bibitem{qsr2014}
QSR International. (2014). \textit{NVivo 10: Getting Started Guide}.

\bibitem{sampaio2018}
Sampaio, R. C., \& Lycarião, D. (2021). \textit{Análise de conteúdo categorial: manual de aplicação}. Brasília: Enap.

\bibitem{sousa2020}
Sousa, J. R., \& Santos, S. C. M. (2020). Análise de conteúdo em pesquisa qualitativa: modo de pensar e de fazer. \textit{Pesquisa e Debate em Educação}, 10(2), 1396-1416.

\bibitem{zhou2020}
Zhou, M. et al. (2020). Progress in neural NLP: Modeling, learning, and reasoning. \textit{Engineering}, 6(3), 275-290.

\end{thebibliography}

\clearpage
\appendix
\section{Prompts Used in the System}

The following prompts were engineered to guide the LLM model (Gemini) through the specific methodological steps of Bardin's Content Analysis.
\subsection{Prompt: Pre-Analysis Stage}
\begin{lstlisting}
You are at the pre-analysis stage and will carry out the identification of indices and the selection of indicators from the analysis documents, following Laurence Bardin's methodology.
Based on the analysis documents (corpus), you must:
1. IDENTIFY indices: "Traces" or signs present on the text's surface that suggest the presence of phenomena related to the central question.
2. CONSTRUCT indicators: Define how to systematically measure or identify the presence or absence of these indices.
CRITERIA FOR SELECTING INDICES
Indices must be:
- Pertinent: Related to the central question and hypotheses.
- Observable: Objectively identifiable in the text.
- Manifest: Explicitly present in the discourse.
- Representative: Relevant to the intended inference.
CRITERIA FOR CONSTRUCTING INDICATORS
Indicators must be:
- Precise: Clearly defined and unambiguous.
- Reliable: Allowing another analyst to reach the same results.
- Systematic: Consistently applicable across the entire corpus.
- Fit for purpose: Quantitative or qualitative, as required.
Contextualization documents should be used as sources of theoretical concepts to underpin the analysis.
The central focus of this content analysis is to answer the question driving the analysis the research's central question. This need not necessarily be a question; it could be a thesis or a research motivation.
Proceed with methodological rigor. Your analysis must be replicable by another researcher following the same criteria.
You are at the pre-analysis stage of the methodology and are extracting indices and indicators from the analysis documents, which consist of transcribed interviews. Your response MUST be a single block of JSON code, without Markdown formatting, comments, or introductory text.
Include precise and referenceable information. The JSON must be an object
containing a single key "indices", which is a list of objects.
Additionally, each indicator must include a list of references indicating where it
was sourced from.
Return ONLY a valid JSON object, without markdown, comments, or additional
text:
The JSON MUST FOLLOW THIS STRUCTURE EXACTLY:

{{
"indices": [
{{
"name": "Index Name",
"description": "Clear description of what this index is and

why it was chosen, with theoretical grounding",

"indicator": "PRECISE description of how to measure/identify this index",

"references": [
{{
"document": "file_name",
"page": "2", // Always 1 for TXT documents
"quoted_content": "The exact text excerpt that

justifies this index"
}}
]
}}
]
}}

THE STRUCTURE ABOVE IS THE EXPECTED FINAL STRUCTURE.
You will receive all contextualization data and must populate the
JSON with the indices; each index must contain a name and a description. The
description of the index should be brief and include a comment regarding the
choice of the index and/or what it represents in the analysis.
CENTRAL RESEARCH QUESTION:
{0}
CONTEXTUALIZATION DOCUMENTS SUBMITTED (names):
{1}
ANALYSIS DOCUMENTS SUBMITTED (names):
{2}
CONCRETE EXAMPLES OF INDICES AND INDICATORS (fictitious context:
Analysis of interviews with teachers about their experiences during the
pandemic):
{{
"indices": [
{{
"name": "Technological difficulties",
"description": "Mentions of problems with the internet,

platforms, or computers",

"indicator": "Presence of words: 'internet went down', 'didn't

work', 'froze', 'couldn't access'"

}},
{{
"name": "Fatigue and exhaustion",
"description": "Expressions related to physical

or mental fatigue",

"indicator": "Presence of words: 'tired', 'exhausted',

'drained', 'can't take it anymore'"

}},
{{
"name": "Social isolation",
"description": "Mentions of a lack of human contact",
"indicator": "Presence of words: 'alone', 'lack of

contact', 'distant', 'miss the students'"

}},
{{ 
"name": "Learning new tools",
"description": "Accounts regarding learning to use

technologies",

"indicator": "Presence of expressions: 'learned to use',

'discovered', 'mastered', 'gained skills'"

}},
{{
"name": "Schedule flexibility",
"description": "Mentions of time-related

benefits",

"indicator": "Presence of expressions: 'working from

home', 'my schedule', 'more time', 'avoided traffic'"

}}
]
}}

ADDITIONAL THEORETICAL FRAMEWORK REGARDING THE COLLECTION OF INDICES AND
INDICATORS (excerpt taken from Bardin's book):
NOTE: THIS CONTENT IS EXCLUSIVELY FOR CONTEXTUALIZING THE STAGE

YOU ARE CURRENTLY UNDERTAKING

If texts are considered a manifestation containing indices that the analysis will make explicit, the preparatory work involves selecting them based on hypotheses, should they be established and systematically organizing them into indicators.

As another example, the index might be the explicit mention of a theme in a message. If one proceeds from the premise that a theme holds greater importance for the speaker the more frequently it is repeated (as in quantitative systematic analysis), the corresponding indicator will be the frequency of that theme either relative or absolute in relation to others.
For example: it is assumed that emotion and anxiety manifest
through speech disturbances during a therapeutic interview. The selected
indices

("uh-huh," interrupted phrases, repetition, stuttering, incoherent

sounds...) and

their frequency of occurrence will serve as an indicator of the underlying

emotional state.

Once the indices have been chosen, the next step is to construct

precise and reliable

indicators. From the pre-analysis stage onward, one must determine

procedures for

segmenting the text into comparable categorization units for

thematic analysis and for establishing coding methods for data recording.

Generally, the effectiveness and relevance of the

indicators are verified by testing them on a few passages or elements of

the documents (pre-test analysis).

[...]
Only indices are retained in a non-frequency-based manner, allowing the analyst to resort to quantitative tests for example, the appearance of similar indices in similar discourses.
[...]
The aim of content analysis is to infer knowledge regarding the conditions of production (or, potentially, reception) an inference

that relies on indicators (whether quantitative or not).
[...]
Let us consider an example: I intend to measure a subject's level of anxiety which they do not consciously express in the message they have transmitted a task that requires, *a posteriori*, a written transcription of the spoken word and various manipulations. I might decide to adopt an indicator of a semantic nature. For example (at the level of meaning), noting the frequency of terms or themes related to anxiety within the subject's vocabulary. Or else, if it seems valid to me, I might use a linguistic indicator (the sequence of signifiers, the length of "sentences"), or a paralinguistic one (intonation and pauses).
Ultimately, the scope, functioning, and objective of content analysis can be summarized as follows: currently, and in general terms, the term "content analysis" refers to:

 A set of communication analysis techniques aimed at obtaining through systematic and objective procedures for describing message content indicators (quantitative or not) that allow for the inference of knowledge regarding the conditions of production/reception (inferred variables) of those messages.


THEORETICAL SUMMARY OF THE STAGE:

Furthermore, during this phase, indicators are developed to underpin
the final interpretation (BARDIN, 2011). If we consider that the documents
(the corpus) are a manifestation containing "traces" or indices that the
analysis will make "speak," the mission of this stage is to select these indices and
organize them into indicators in a systematic way.

Indices (traces) are the signs on the surface of the text or
communication that suggest the presence of something you wish to investigate. Indicators (measures),
on the other hand, are the way you will measure or quantify (or
not) the presence of these indices so that you can then infer knowledge
about the context (the inferred variable).

Imagine your hypothesis is: The patient's emotion and anxiety
manifest when they speak during a therapeutic interview.
In this example, you transform the abstract phenomenon ("anxiety") into a
concrete indicator (the "frequency count" of incoherent sounds),

77

allowing the analysis to be objective and systematic, going beyond what the
patient explicitly said.

Concept to be Inferred (Inferred Variable) = Anxiety/Emotion
Hypothesis = Anxiety manifests through speech disturbances
Indices (Traces in the Text) = Interrupted sentences, repetition

of words, stuttering, incoherent sounds (e.g., "uh," "um")

Indicator (Systematic Measure) = The frequency of occurrence

of these indices
\end{lstlisting}
\clearpage
\subsection{Prompt: Exploration Stage}
\begin{lstlisting}
You are at the MATERIAL EXPLORATION stage and will be grouping the **units of analysis** into **categories**. 
The central focus of this content analysis is to answer the question driving the analysis the research's central question. It may not necessarily be a question, but rather a thesis or a research motivation.
Proceed with methodological rigor. Your analysis must be replicable by another researcher following the same criteria. Your task: Based on the provided indices and indicators, you must:
1. IDENTIFY RECORDING UNITS in the analysis documents that contain
the defined indices
2. GROUP these recording units into thematic CATEGORIES
3. COUNT the frequency of each category (how many recording units
belong to it)
COUNTING CRITERIA (Unit of Enumeration)
Type: Simple frequency (as described by Bardin)
Recording unit: Text segment (sentence/excerpt) expressing a
complete idea
Referent: Only statements made by the interviewees
Counting rule: Count +1 for each DISTINCT recording unit that:
1. Contains at least one of the provided indices/indicators
2. Semantically expresses the category's content
3. Is not a repetition of the same theme within the same context
4. Has a literal meaning (disregard irony or negations)
STRICT RULES
1. Use ONLY the provided indices and indicators (refer to them by name
whenever **citing them**, never by their IDs)
2. Each recording unit must cite the EXACT excerpt from the source document
3. Categories must emerge from the thematic grouping of the units
4. Do not invent or infer information not present in the texts
5. Justify each classification based on the indices found
6. Adhere faithfully to Bardin's method: categories must be mutually
exclusive, homogeneous, pertinent, objective, and productive
IMPORTANT NOTE:
There is NOT necessarily a 1:1 relationship between indices and categories! - Indices are concrete elements in the text (words, phrases, expressions)
- Recording units are text segments containing one or more indices
- Categories are thematic groupings that emerge from multiple recording
units
Return ONLY a valid JSON object, without markdown, comments, or additional
text:
{{
"categories": [
{{
"name": "Category Name",
"definition": "Clear description of the semantic criterion defining this
category",
"frequency": 0,
"register_units": [
{{
"text": "EXACT excerpt taken from the document",
"document": "file_name",
"page": "2", //always 1 for TXT documents
"found_indices": ["index_id_1", "index_id_2"],
"indicator": "indicator_id_X",
"justification": "Explanation of why this unit belongs to the
category"
}}
]
}}
]
}}
You will receive all the contextualization data and must fill in the
JSON with the categories.
The definitions of Laurence Bardin's methodology must be followed strictly.
THERE IS NOT NECESSARILY A 1:1 RELATIONSHIP BETWEEN CATEGORY AND INDEX.
THEY ARE METHODOLOGICALLY DIFFERENT THINGS. EXAMPLE OF INDEX-BASED CATEGORIES (fictitious context: Analysis of
interviews with teachers regarding their experiences during the pandemic):
{{
EXAMPLE OF POSSIBLE INDICES AND INDICATORS FOR AN ANALYSIS:
"indices": [
{{
"id": "idx_001",
"name": "Technological difficulties",
"description": "Mentions of problems with the internet, platforms,
or computers",
"indicator": "Presence of words: 'internet went down', 'didn't work',
'froze', 'couldn't access'"
}},
{{
"id": "idx_002",
"name": "Fatigue and exhaustion",
"description": "Expressions related to physical or mental fatigue",
"indicator": "Presence of words: 'tired', 'exhausted', 'drained',
'can't take it anymore'"
}},
{{
"id": "idx_003",
"name": "Social isolation",
"description": "Mentions of a lack of human contact",
"indicator": "Presence of words: 'alone', 'lack of contact',
'distant', 'miss the students'"
}},
{{
"id": "idx_004",
"name": "Learning new tools",
"description": "Accounts of learning to use technologies",
"indicator": "Presence of expressions: 'learned to use', 'discovered',
'mastered', 'got trained'"
}},
{{
"id": "idx_005",
"name": "Schedule flexibility",
"description": "Mentions of time-related benefits",
"indicator": "Presence of expressions: 'work from home', 'my
schedule', 'more time', 'avoided traffic'"
}}
]
}}
POSSIBLE CATEGORIES FOR THE INDICES AND INDICATORS LISTED ABOVE:
{{
"categories": [
{{
"name": "Remote Teaching Challenges",
"definition": "Groups difficulties and obstacles faced during
online teaching",
"frequency": 15,
"register_units": [
{{
"text": "The internet went down three times during the class; it
was horrible",
"document": "entrevista_prof_01.pdf",
"page": "2",
"found_indices": ["idx_001"],
"main_indicator": "idx_001",
"justification": "Reports a technical issue that disrupted the class"
}},
{{
"text": "I felt completely drained; I didn't have the energy to
prepare lessons",
"document": "entrevista_prof_01.pdf",
"page": "3",
"found_indices": ["idx_002"],
"main_indicator": "idx_002",
"justification": "Expresses exhaustion related to remote work"
}},
{{
"text": "Zoom froze, and I lost all the students in the room",
"document": "entrevista_prof_03.pdf",
"page": "1",
"found_indices": ["idx_001"],
"main_indicator": "idx_001",
"justification": "Another technical issue that impacted the class"
}},
{{
"text": "Alone in front of the screen, without seeing anyone's face,
I felt isolated",
"document": "entrevista_prof_02.pdf",
"page": "4",
"found_indices": ["idx_003"],
"main_indicator": "idx_003",
"justification": "Reports the emotional impact of isolation" 
}}
]
}},
{{
"name": "Gains and Lessons Learned",
"definition": "Groups positive aspects and benefits identified
in the experience",
"frequency": 8,
"register_units": [
{{
"text": "I learned to use Moodle, Google Classroom, and various
new tools",
"document": "entrevista_prof_02.pdf",
"page": "2",
"found_indices": ["idx_004"],
"main_indicator": "idx_004",
"justification": "Reports the acquisition of technological
skills"
}},
{{
"text": "Working from home, I had more time with my family",
"document": "entrevista_prof_04.pdf",
"page": "3",
"found_indices": ["idx_005"],
"main_indicator": "idx_005",
"justification": "Identifies a benefit related to
flexibility"
}},
{{
"text": "I trained in digital tools I had never
used before",
"document": "entrevista_prof_01.pdf",
"page": "5",
"found_indices": ["idx_004"],
"main_indicator": "idx_004",
"justification": "Another report of skills
development"
}},
{{
"text": "Avoiding two hours of traffic a day was liberating",
"document": "entrevista_prof_03.pdf",
"page": "2",
"found_indices": ["idx_005"],
"main_indicator": "idx_005",
"justification": "Benefit related to time saved"
}}
]
}}
]
}}
}}
INDICES AND INDICATORS (RESULTING FROM THE PRE-ANALYSIS STAGE):
{0}
CENTRAL RESEARCH QUESTION:
{1}
CONTEXTUALIZATION DOCUMENTS SUBMITTED (names):
{2}
ANALYSIS DOCUMENTS SUBMITTED (names):
{3}
ADDITIONAL THEORETICAL FRAMEWORK REGARDING THE MATERIAL EXPLORATION AND
CODING STAGE (excerpts taken from Laurence Bardin's book):
NOTE: THIS CONTENT IS EXCLUSIVELY FOR CONTEXTUALIZING THE STAGE
YOU ARE CURRENTLY UNDERTAKING
If the various pre-analysis operations are properly
completed, the analysis phase itself is nothing more than the
systematic application
of the decisions made. Whether involving procedures applied
manually or operations performed by computer, the execution of the
program proceeds mechanically. This phase, long and tedious, consists

essentially of coding, decomposition, or enumeration operations,

based on previously formulated rules.

[...]

Processing the material means coding it. Coding corresponds to a transformation carried out according to precise rules of the raw text data, a transformation that, through segmentation, aggregation, and enumeration, allows one to arrive at a representation of the content or its expression; one capable of enlightening the analyst regarding the text's characteristics,
which can serve as indices, or, as O. R. Holsti puts it: Coding is the process by which raw data are systematically transformed and aggregated into units, which allow for an exact description of the pertinent characteristics of the content.
[...]
The organization of coding involves three choices (in the case of a
quantitative and categorical analysis):
* Segmentation: choice of units;
* Enumeration: choice of counting rules; (NOTE: THIS STAGE HAS ALREADY
BEEN COMPLETED AND WILL BE PROVIDED)
* Classification and aggregation: choice of categories. (NOTE: WE ARE
CURRENTLY AT THIS STAGE)
[...]
RECORDING AND CONTEXT UNITS
Which elements of the text should be taken into account? How should the text be segmented into complete units? The choice of recording and context units must be made appropriately (appropriateness regarding the material's characteristics and the analysis objectives).

a) The recording unit: This is the coded unit of meaning; it corresponds to the content segment treated as the basic unit for the purposes of categorization and frequency counting. The recording unit can vary widely in nature and size. 
A certain ambiguity prevails regarding the criteria for distinguishing recording units. Indeed, certain segmentations are carried out at a semantic level for example, the "theme" while others are made at an apparently linguistic level, such as the "word" or the "sentence."
This serves as a critique of disciplines whose scientific and rigorous nature is more evident. 
In fact, the segmentation criterion in content analysis is always semantic in nature, even though there may sometimes be a correspondence with formal units (examples: word and keyword/thematic word; sentence and signifying unit).
By way of illustration, the following can be cited among the most commonly used recording units:

* The word: while it is true that the "word" lacks a precise definition in linguistics, for language users it corresponds to something tangible. However, linguistic precision can be introduced if relevant. 

All words in the text may be taken into account, or one may retain only keywords or thematic words (*symbols* in English); a distinction can also be made between content words and function words;
or an analysis of a specific word category may be performed: nouns, adjectives, verbs, adverbs (...) in order to establish ratios

[...]
The context unit:
The context unit serves as the unit of comprehension for coding the recording unit; it corresponds to the segment of the message whose dimensions (larger than those of the recording unit) are optimal for grasping the exact meaning of the recording unit.
This might, for example, be the sentence for a word and the paragraph for a theme.
Indeed, in many cases, it becomes necessary to refer (consciously) to the immediate or broader context of the unit being recorded. If multiple coders are working on the same corpus, prior agreement is essential.
For instance, when analyzing political messages, words such as liberty, order, progress, democracy, and society require context to be understood in their true sense.

Reference to context is crucial for both evaluative analysis and contingency analysis. Results are liable to vary significantly depending on the dimensions of the context unit. The intensity and scope of a unit may appear more or less pronounced, depending on the dimensions of the chosen context unit. 
Regarding co-occurrences, it is evident that their number increases with the dimensions of the context unit: for example, it is unlikely that similar themes would be found within a single paragraph or a few minutes of a recording,
whereas the probability increases in a multi-page text or a one hour broadcast. Generally, the larger the context unit, the more strongly attitudes or values emerge in an evaluative analysis, or the more numerous the co-occurrences in a contingency analysis.
Determining the dimensions of the context unit is governed by
two criteria: cost and relevance. Clearly, a unit of broad context requires a more extensive re-examination of the material. On the other hand, there is an optimal dimension in terms of meaning: if the context unit is too small or too large, it is no longer suitable; here, too, both the type of material and the theoretical framework are decisive factors.

In any case, it is possible to test the recording and context units
on small samples to ensure that we are working with the
most appropriate instruments
[...]
138 CONTENT ANALYSIS
2. ENUMERATION RULES
A distinction must be made between the recording unit what is counted and the enumeration rule the method of counting.
Let us consider the following example: we have a completed "text"
in which the identification and segmentation processes yielded the following
elements or recording units (words, themes, or other units):
a, d, a, e, a, b. Given that the reference list established based on a set of "texts" or according to a standard is a, b, c, d, e, f, it is possible to use various types of enumerations:

Frequency (NOTE: THE MEASURE TO BE USED IN OUR CASE): this is generally the

most widely used measure. It corresponds to the

following postulate (valid in some cases but not others): the importance of

a recording unit increases with its frequency of occurrence. In our

example, the frequency of each element is:
a = 3;
b = 1;
c = 0;
d = 1;
e = 1;
f = 0.

A frequency measure in which all occurrences carry the same weight

postulates that all elements are of equal importance.

The choice of a simple frequency measure should not be automatic. We must

remember that it rests on the following implicit assumption: the occurrence of an item of meaning or expression becomes increasingly significant in relation to the intended description or interpretation of the reality under study the more frequently it recurs. Quantitative

regularity of occurrence is, therefore, what is deemed significant. This

presupposes that all items have the same value, which is not

always the case.
[...]

Categorization is an operation involving the classification of the constituent

elements of a set through differentiation and subsequent regrouping

according to genre (analogy), based on

predefined criteria.
Categories are headings or classes that gather a group of elements (recording units, in the case of content analysis) under a generic title - a grouping performed based on the common characteristics of these elements.
The categorization criterion may be semantic (thematic categories: for example, all themes signifying anxiety are grouped under the category "anxiety," while those signifying relaxation are grouped under the conceptual heading "relaxation"), syntactic (verbs, adjectives),
lexical (classification of words according to their meaning, pairing synonyms and closely related meanings), and expressive (for example, categories that
classify various language disorders).
[...]
CONTENT ANALYSIS
Taxonomic activity is a very common operation involving the distribution
of objects into categories. If, before placing a record on the turntable,
we ask ourselves whether we want to listen to Bach, Ravel, or Boulez,
we are not using the same criterion that guides our choices if we ask ourselves
whether we wish to hear the violin, organ, or piano. The categorization criterion
is not the same (composer versus instrument). We are not emphasizing the same aspect of reality. Furthermore, the criterion we employ is more or less suited to the reality presented to us. It is possible that our two desires might converge, thereby refining the choice we make (a specific instrument and a specific composer). Similarly, in content analysis, a message may be subjected to one or several dimensions of analysis.

Classifying elements into categories requires investigating what each one has in common with the others. What enables their grouping is the common element shared among them. However, other criteria might focus on different aspects of similarity, potentially altering the previous distribution significantly.

Categorization is a structuralist process comprising two
stages:
* inventory: isolating the elements;
* classification: distributing elements and, consequently, seeking or imposing
a certain organization on the messages.
[...]
EXAMPLES OF CATEGORY SETS

While in most cases it is necessary to create a category grid for each new analysis, previous studies can serve as inspiration for the analyst. For this reason, we will cite some examples of category sets that have already been used.

CATEGORIZATION
a) Value analysis
Example 1:
Shortly after World War II, White specialized in value analysis.
He first analyzes Richard Wright's autobiography, *Black Boy* (1947);
he then analyzes the propaganda styles of Hitler and Roosevelt (1949)
and, later, the speeches of Kennedy and Khrushchev (1967).

We present one of his analysis grids. A / Physiological values
1. Food/Nutrition
2. Sex
3. Rest
4. Health
5. Safety/Security
6. Comfort
D / Values expressing fear
(emotional security)
E / Values of play and joy
1. New experience
2. Excitement, emotion
3. Beauty
4. Humor
5. Creative self-expression
F / Practical values
1. Practicality
2. Possession
3. Work
G / Cognitive values
1. Knowledge
H / Miscellaneous
1. Happiness
2. Value in general
B / Social values
1. Sexual love
2. Family love
3. Friendship
C / Ego-related values
1. Independence
2. Achievement/Fulfillment
3. Recognition
4. Self-esteem/Self-love
5. Domination
6. Aggression
Example 2:

V. Isambert-Jamati demonstrated the evolution of values advocated by the
school system between 1860 and 1965, based on the analysis of a sample of prize-giving speeches elivered by various speakers directly or indirectly involved in secondary education, produced regularly during 2. R. K. White, Value-analysis: the nature and use of the method, Glen Gardiner, N. J., Libertarian Press, 1951. Cited by Holsti, op. cit.

3. Isambert Jamati, Crises de la societe, crises de l'enseignement, P. U. F., 1970.

CONTENT ANALYSIS
that period and easily accessible. These prize-giving speeches served as
source material for a comprehensive study on the school's "reference morality," regarding the ends and the means to achieve those ends pursued by the school system, as well as the objects of intellectual knowledge to be promoted, etc.

A set of five categories and subcategories served as the basis for the analysis.
* The changes that the teaching of school subjects should bring about in
students:
* Participation in supreme values.
* Individual self-improvement sought by the student.
* Exercise of operational mechanisms.
* The objects of knowledge:
* People of the past and their works.
* Contemporary people.
* Human and universal nature.
* Nature.
* The objects of moral education:
* Loyalty to the secular national university,
* Loyalty to the institution.
* Withdrawal from the world as a favorable condition for education,
* Educational value of discipline.
* Role of peers in character formation.
* Consideration of individual differences among students.
* Utilization of playful tendencies.
* Teachers' moral example.
* Teachers' voluntary influence.
* The institutional definition:
* It is beneficial for the core definition of secondary education to change
so that it adapts to social shifts.
* Secondary schooling should be of long duration.
* Secondary education should suffice for students, without the need
for them to continue their studies.
* Lycees should not serve to prepare students for their future careers.
* The target audience is the social elite.

CATEGORIZATION
* Reference values:
* Individual morality based on perfection or the categorical imperative. 
* Individual morality with a hedonistic tendency, or of the "mental hygiene" type:
* Individual morality based on solidarity.
* Exhortation to work.
* Exaltation of progress.
* Exaltation of youth.
* Exaltation of the family.
* Exaltation of the homeland.
* Exaltation of peace and international understanding.
The final conclusion of this study demonstrates that changes in French society
are reflected in the objectives proposed by educational systems, and
that societal crises and educational crises appear synchronized. The objectives of the school institution evolve. Thus, it is possible to divide the periods according to dominant values:

1) 1860-1870: supreme values and integration into the elite.
2) 1876-1885: integration into the elite and transformation of the world.
3) 1896-1905: transformation of the world and secular enthusiasm.
4) 1906-1930: culture as a free good.
5) 1931-1940: learning to learn.
6) 1946-1960: secondary education defends itself: return to aestheticism.
7) 1961-1965: crisis of objectives.

From a technical standpoint, the analyses were essentially thematic, yet always refined by precautions such as the weighting of themes, the division into primary and secondary themes, an evaluative approach (favorable text vs. neutral text), and the use of gender relation specifically a "dominance coefficient":

b) Analysis of ends and means
Example 1:

This is an analysis of affective and rational objectives conducted by B. Berelson and P. Salter regarding popular fiction magazines.

Two category systems were used:

4. And then came May '68! 5 B. Berelson and P. Salter, "Majority and minority Americans: an analysis of magazine fiction," *Public Opinion Quarterly*, 1946.

CONTENT ANALYSIS

A / Intentions of the "heart"
1. Romantic love
2. Established marriage
3. Idealism
B / Intentions of the "head"
1. Solving concrete problems
2. Personal progress
3. Money and material goods
4. Affection and emotional security
5. Economic and social security
6. Power and domination
7. Patriotism
8. Adventure
9. Justice
10. Independence
Example 2:
This study analyzes the goals and possibilities for success offered to children in television programs, relating them to the advocated means.

A / Goal categories
1. Property (material success)
2. Self-preservation (including desire for the status quo)
3. Affection
4. Sentiment
5. Power and prestige
6. Psychological objectives (including violence and education)
7. Others
B / Method categories
1. Legal
2. Non-legal (without injury or damage)
3. Economic
4. Violence
5. Organization, negotiation, and compromise
6. Evasion, flight (attempt to avoid facts inherent to achieving the
goal, forgetting the objective, etc.)
7. Chance
8. Others
6. O. N. Larson, L. N. Gray, and J. G. Fortis, "Goals and goal-achievement
methods in television
content: model for anomie?" in *Social Inquiry*, 33, 1963.

\end{lstlisting}

\end{document}